\documentclass[aps,
floatfix,superscriptaddress,a4paper]{revtex4}
\usepackage{graphicx, amsmath,mathrsfs,amssymb,psfrag}
\usepackage{bbm}
\usepackage{enumerate}

\newcounter{enumi_saved}
\newenvironment{myenumerate}{\begin{enumerate}\setcounter{enumi}{\value{enumi_saved}}}
{\setcounter{enumi_saved}{\value{enumi}}\end{enumerate}}

\evensidemargin \oddsidemargin
\begin{document}

\author{Laure Gouba}
\affiliation{
The Abdus Salam International Centre for
Theoretical Physics (ICTP),\\
 Strada Costiera 11,
I-34151 Trieste Italy. \\
Email: lgouba@ictp.it}

\title{A comparative review of four formulations of noncommutative quantum mechanics}

\begin{abstract}
 Four formulations of quantum mechanics on noncommutative Moyal phase spaces are reviewed. These are 
the canonical, path-integral, Weyl-Wigner and systematic formulations. Although these formulations all 
represent quantum mechanics  on a phase space with the same deformed Heisenberg algebra, 
there are mathematical and conceptual differences which we discuss.\\


\end{abstract}

\maketitle

\section{Introduction}

Quantum mechanics is now more than a hundred years old, starting from
the Max Planck's description in $1900$ of the black body radiation problem 
and the paper of Albert Einstein in $1905$ about quantum-based theory 
to explain the photoelectric effect. However quantum theory has been reformulated 
around $1925$ with the fundamental Heisenberg commutation relations or phase space 
algebra and the associated uncertainty principle. One of the fundamental features is to deal 
with operators that do not commute, more precisely, the commutations relations 
between positions and momenta. The terminology noncommutative quantum mechanics is 
a terminology to express that the positions operators do not commute, more generally as we will review below,
one can consider the situation where the commutators for coordinates and momenta are non-canonical.

One of the problems in the $1930$s was how to resolve infinities in the then newly 
introduced quantum field theory. 
The idea of extending noncommutativity to the coordinates as a possible way of removing 
the infinite quantities appearing in field theories was first suggested by Werner Heisenberg. 
That is the birth of the idea of quantum spacetime, which is a generalization of the usual 
concept of spacetime in which some variables that ordinarily commute are assumed to be noncommuting
and form a different Lie algebra. 

He passed the idea to his doctoral student Rudolf Peierls in the late $1930$s \cite{peierls}.
Peierls made use of these ideas eventually in his work related to the Landau level problem. He 
also noticed that electrons in a magnetic field can be regarded as moving in a quantum space.
He also passed the idea to Wolfgang Pauli who then involved Robert Oppenheimer in the discussion \cite{pauli}.
Oppenheimer carried it to his student Hartland Snyder, who published the first concrete example in 1947
\cite{snyder}. It was a period when ideas about renormalization also born and the success of the 
renormalization theory took over the ideas about noncommutative coordinates.
Later in $1980$s - $1990$s Alain Connes developped noncommutative geometry \cite{connea,conneb} and that 
has been very successfull and plausible that it has attracted the attention of particle physicists 
and string theorists
\cite{wittene, doplicher, aconne, sjabbari, seiberg, mmjabbari, douglas, bich, szabo}.
The history about noncommutative spacetime can be also found in \cite{bal}. 

The simplest noncommutativity one can postulate is that the space-time coordinates satisfy the commutation relation 
\begin{equation}\label{ecua}
 [\hat x_i,\hat x_j] = i\theta_{ij},
\end{equation}
where $\theta_{ij}$ is a constant anti-symmetric tensor of dimension $(length)^2$. Some of the consequences of 
the space-time coordinates noncommutativity are the following:
\begin{itemize}
\item the action for quantum field theory on noncommutative space is obtained from a 
usual quantum field theory through replacing the field products by 
Moyal star-product \cite{filk};
\item an uncertainty relation between position measurements will a priori lead to a nonlocal theory; 
\item the presence of the constant $\theta_{ij}$ in equation (\ref{ecua}) violates Lorentz invariance 
(if the dimension of space-time is greater than two); 
\item there is a persistence of the ultraviolet divergences 
in noncommutative quantum field theory \cite{douglas,szabo}; 
\item the time, space noncommutativity leads to violation 
of unitarity and causality \cite{susskind}.
\end{itemize}

Despite the above (perhaps unwanted ) features, there are strong motivations in studying 
noncommutative space-time.  There is a long-held belief that in quantum theories including gravity,
space-time must change its nature at distances comparable to the Planck scale. 
One might also study noncommutative theories 
as interesting analogs of theories of more direct interest, such as Yang-Mills theory. 
It was found that the equation (\ref{ecua}) follows naturally as a particular low-energy limit 
of string theories with $\theta_{ij}$ directly related to a constant antisymmetric background field $B_{ij}$ 
in the presence of D-brane \cite{seiberg, mmjabbari}. Noncommutative field theory is also known to appear naturally 
in condensed-matter theory, for instance the theory of electrons in a magnetic field projected to the lowest Landau level 
(the Quantum Hall problem), which is naturally thought of as a noncommutative Chern-Simons theory \cite{lsuskind}.

In this work, however, we focus on quantum mechanics on noncommutative spaces, rather than quantum field theory.
Since in noncommutative quantum mechanics with commutative  time evolution, we do not encounter the problems 
 that may occur in noncommutative quantum field theory,  it is
well suited as a toy model for the introduction of noncommuting coordinates.  Indeed, 
noncommutative quantum mechanics has drawn attention of many authors, 
\cite{horvata,horvatb,horvatc,horvatd,
gamboaja,gamboajb,gamboajc,gamboajd,
polya,polyb,
sbelucia,sbelucib,sbelucic,
smailagicaa,smailagicab,smailagicac,
chachaa,chachab,deri,meljanaca,meljanacb,
scholtza,scholtzb,scholtzc,scholtzd,
alavia,alavib,
bertolamioa,bertolamiob,bertolamioc,bertolamiod,
lia,lib,lic,lid,
girottia,girottib,girottic,
ali,subir,dey,hounk}. We would like to apologize for not beeing 
able to cite all the papers relating the 
topic. In 2008, we have been interested in the formulation and 
interpretation of noncommutative quantum mechanics \cite{fgs} and that led us to some significant contributions 
in the development of noncommutative quantum mechanics 
\cite{laurea,laureb,laurec,laured,lauree,lauref,laureg,laureh,laurei}.

Various formulations of noncommutative quantum mechanics are found in the literature. Are they 
equivalent\;? Do they provide different insights\;? Are there some problems more difficult in one formulation 
and more easy in another one\;? In this paper, we attempt to give a catalogue of formulations of noncommutative 
quantum mechanics. Our intention is primarily to examine the distinction between these mathematical formulations 
and briefly discuss then the bearing this may have on the conceptual interpretations.

\section{Catalogue of formulations}

\subsection{Canonical formulation}

\subsubsection{New coordinates system}
The noncommutative space is realized by the coordinates operators satisfying 
 \begin{equation}
  [ \hat x_i, \hat x_j ] = i\theta_{ij}, 
 \end{equation}
where the operators $\hat x_i,\; i =1,2$ are the coordinate operators and $\theta_{ij}$ is real and 
antisymmetric, i.e., $\theta_{ij} = \theta\epsilon_{ij}$, with $\epsilon_{ij}$ the completely antisymmetric 
tensor $\epsilon_{12} = 1$ and the spacial noncommutative parameter $\theta$ is of dimension of $(length)^2$.
 The choice of the phase space is such that
\begin{equation}\label{eq1}
 [\hat x_i,\;\hat x_j] = i\theta_{ij}; \quad [\hat x_i,\hat p_j] = i\hbar\delta_{ij};\quad [\hat p_i,\hat p_j] =0,\quad 
 i,j =1,2,
\end{equation}
or in a general setting where the momenta are also noncommutative we consider
\begin{equation}\label{eq2}
 [\hat x_i,\;\hat x_j] = i\theta_{ij}; \quad [\hat x_i,\hat p_j] = i\hbar\delta_{ij};\quad [\hat p_i,\hat p_j] = i\eta_{ij},\quad 
 i,j =1,2\;,
\end{equation}
where $\eta_{ij} = \eta\epsilon_{ij}$ is totally antisymmetric representing the noncommutativity property of the momentum on 
noncommutative phase space. 

The approach here is to express the set of noncommutative coordinates as linear combination 
of the canonical variables or vice versa. 
In the situation of equation (\ref{eq1}), the most used, 
the Hilbert space can be consistently be taken to be exactly the same as the Hilbert space of the corresponding commutative system.
The Hamiltonian 
has to be introduced, and that is nontrivial. Once it is done, the dynamical equation for the state $\vert\psi\rangle$ is the usual 
Schr\"odinger equation, $H\vert \psi\rangle = i\hbar\frac{\partial}{\partial_t}\vert \psi\rangle$. 

A new coordinates system is introduced as follows
\begin{equation}\label{co1}
  x_i = \hat x_i + \frac{1}{2\hbar}\theta_{i,j} p_j,\quad  p_i = \hat p_i, 
\end{equation}
where the new variables satisfy the usual canonical commutation relations 
\begin{equation}\label{eq3}
 [x_i,\;x_j] = 0;\quad [x_i,\;p_j] = i\hbar\delta_{ij};\quad [p_i,p_j] = 0.
\end{equation}
Since the noncommutativity parameter $\theta$ is very small, the 
noncommutativity effects can be treated as some perturbations of the commutative counterpart up to first 
order $\theta$, then one can use the usual wave functions and probabilities. The difference between noncommutative 
and ordinary quantum mechanics consists in the choice of polarisation. 

In the situation of equation (\ref{eq2}), the representation of $\hat x_i$ and $\hat p_i$,  could be the following
\begin{equation}
 \hat x_i = \kappa x_i - \frac{1}{2\hbar\kappa}\theta_{ij}p_j\;, \quad \hat p_i = \kappa p_i + \frac{1}{2\hbar\kappa}\eta_{ij} x_j\;, 
 \quad i, j = 1,2,
\end{equation}
where $\kappa$ is a scaling constant related to the noncommutativity of phase space. 

{\it Reference papers}:
\begin{myenumerate}
\item M. Chaichian, M.M. Sheikh-Jabbari, A. Tureanu, Hydrogen Atom Spectrum and the Lamb Schift in Noncommutative QED, 
Phys. Rev. Lett.  vol. 86, Num.13 (2001).
\item M. Demetrian, D. Kochan, Quantum Mechanics on non-commutative plane, Acta Physica Slovaca 52, Num 1 pp: 1-9 (2002).
\item S. Bellucci, A. Nersessian, C. Sochichiu, Two phases of the noncommutative quantum mechanics, 
Phys. Lett. B 52: 345-349, (2001).
\item  Anais Smailagic, Euro Spallucci, Isotropic representation of the noncommutative $2D$ harmonic oscillator, 
Physical Review D, Volume $65$, 107701, (2002).

\subsubsection{Bopp's shift and star products}

The physicist Fritz Bopp was the first to consider in his paper, where he discussed 
some statistical implications of quantization, pseudo-differential operators obtained from a 
symbol by the quantization rules
\begin{equation}
 x  \rightarrow  x + \frac{1}{2}i\hbar\partial_p\;,\quad 
 p \rightarrow  p-\frac{1}{2} i\hbar\partial_x\;,
\end{equation}
instead of the usual correspondance $x\rightarrow x\;,\: p\rightarrow -\frac{1}{2}i\hbar\partial_x$  \cite{bopp}.
In the physics literature, the operators $ x  \rightarrow  x + \frac{1}{2}i\hbar\partial_p$ and 
$p \rightarrow  p-\frac{1}{2} i\hbar\partial_x $ are called Bopp's shifts and this quantization procedure 
is called Bopp quantization. There is a connection between 
the Bopp's shifts and the star product or $\star$-product, that is an associative deformation of 
ordinary products on phase space. The $\star$-product has been defined 
by Groenewold \cite{groe} as follows
\begin{equation}
 \star \equiv e^{\frac{i\hbar}{2}(\stackrel{\leftarrow}{\partial_x}\stackrel{\rightarrow}{\partial_p} 
 -\stackrel{\leftarrow}{\partial_p}\stackrel{\rightarrow}{\partial_x})}.
\end{equation}
The $\star$-product induces Bopp's shifts \cite{curt} in the sense that it may be evaluated through translations 
of functions arguments 
\begin{equation}
 f(x,p)\star g(x,p) = f(x+ \frac{i\hbar}{2}\stackrel{\rightarrow}{\partial_p}\;,\; p-\frac{i\hbar}{2}
 \stackrel{\rightarrow}{\partial_x})g(x,p).
\end{equation}
In this approach noncommutative quantum mechanics is considered as a theory defined on a manifold where the product 
of functions is the Moyal one \cite{moyal}. If $f(x) $ and $g(x)$ are two functions, then the Moyal product is defined 
as 
\begin{equation}
 (f \star g) (x) = e^{\frac{i}{2}\theta_{ij}\partial_i^{(1)}\partial_j^{(2)}}f(x_1)g(x_2)\vert_{x_1 =x_2 = x}.
\end{equation}
In this sense, for instance, the time dependent Schr\"odinger equation 
\begin{equation}
 i\frac{\partial\psi(\mathbf{x},t)}{\partial t} = \left(\frac{\mathbf{p}^2}{2m} + V(\mathbf{x})\right)\psi(\mathbf{x},t),
\end{equation}
in the noncommutative space is the same one but with the potential shifted as $V(\mathbf{x} -\frac{\tilde{\mathbf{p}}}{2})$, 
where
\begin{equation}
 V(\mathbf{x})\star \psi(\mathbf{x},t) \rightarrow V(\mathbf{x} -\frac{\tilde{\mathbf{p}}}{2})\psi(\mathbf{x},t),
\end{equation}
with $\tilde{p}_i = \theta^{ij}p_j$ and $\theta_{ij} = \theta\epsilon_{ij}$, where 
$\epsilon_{ij}$ is an antisymmetric tensor in two dimensions.
This last fact implies that quantum mechanics in  a noncommutative plane is highly nontrivial because, 
as the shifted potential involves in principle arbitrary powers of the momenta, we will have an arbitrary 
large number of derivatives in the Schr\"odinger equation. 

{\it Reference papers}
\item J. Gamboa, M. Loewe and J. C. Rojas, Non-commutative Quantum Mechanics, Physical Review D,  
Vol. 64. Issue: 6 Article Number: 067901 (2001);
\item
J. Gamboa, M. Loewe, F. Mendez and J. C. Rojas, Noncommutative Quantum Mechanics: The Two-dimensional central Field, 
International Journal of Modern Physics A, Vol. 17 Issue: 19 Pages: 2555-2565 (2002).

The study of noncommutative quantum mechanics can also be reduced to a variant of Bopp calculus.
A variant of Bopp's shift is the map
\begin{equation}\label{bmap}
 \hat x_i = x_i -\frac{1}{2\hbar}\theta_{ij}p_j,\quad \hat p_i = p_i,\; i,j = 1,2,
\end{equation}
that is equivalent to equation (\ref{co1}), where the operators $\hat x_i, \hat p_i, i = 1,2$ satisfy 
the algebra (\ref{eq1}) and the operators $x_i,\;p_i, i=1,2 $ satisfy the algebra (\ref{eq3}). 
In the literature the equation (\ref{bmap}) is called Bopp's shift and this map is linked to 
the Moyal product. 
Let's consider for instance the time independent
the Schro\"dinger equation on noncommutative space 
\begin{equation}
 H(\hat x_i,\hat p_i)\star \psi  = E\psi.
\end{equation}
The Moyal product can be changed into the ordinary product by using the Bopp's shift as
\begin{equation}
  H(\hat x_i,\hat p_i)\star \psi = H (x_i - \frac{1}{2\hbar}\theta_{ij}p_j,\; p_i)\psi = E\psi\;, 
\end{equation}
and the noncommutative effects can be evaluated through the $\theta$ related terms that can be 
treated as a perturbation in ordinary quantum mechanics, since $\theta << 1$. 

{\it Reference papers}:
\item 
Li, K.; Dulat, S., The Aharonov-Bohm effect in noncommutative quantum mechanics
Eur. Phys. J. C Volume: 46 Issue: 3 Pages: 825-828 (2006);
\item
Sahipjamal Dulat, Kang Li, Quantum Hall Effect in Noncommutative Quantum Mechanics, 
Eur. Phys. J.C (2009) 60: 163-168.

\subsubsection{The Seiberg Witten (SW) map}

The Seiberg-Witten (SW) map was discovered by Nathan Seiberg and Edward Witten in the context 
of string theory and noncommutative geometry \cite{seiberg}. They argued that the ordinary 
gauge theory should be gauge equivalent to a noncommutative Yang-Mills field theory. They 
introduced a correspondance between a noncommutative gauge theory and 
a conventional gauge theory.

Let's brieffly review the Seiberg-Witten map, the details being given in \cite{seiberg, bich}. 
Let's consider a flat Minkowski space. On this space, we consider 
the coordinates $x_\mu$ as self-adjoint operators on a Hilbert space, satisfying the algebra
\begin{equation}
 [x_\mu,\; x_\nu] = i\theta_{\mu\nu},
\end{equation}
where $\theta_{\mu\nu}$ is real and antisymmetric. The corresponding field theory is 
equivalent to the usual  commutative flat manifold with the product substituted by 
the non-local $\star$-product 
\begin{equation}
 (f\star g)(x) = \int\frac{d^4k}{(2\pi)^4}
 \int \frac{d^4 p}{(2\pi)^4}e^{-i(k_\mu + p_\mu)x^\mu}
 e^{-\frac{1}{2}\theta^{\mu\nu}k_\mu p_\nu}\tilde{f}(k)\tilde{g}(p),
\end{equation}
where $f$ and $g$ are functions on the manifold. The noncommutative Yang-Mills action is 
\begin{equation}
 \hat \Sigma_{cl} = -\frac{1}{4}\int d^4x\hat F_{\mu\nu}\star \hat F^{\mu\nu} 
 = -\frac{1}{4}\int d^4x\hat F_{\mu\nu}\hat F^{\mu\nu}, 
\end{equation}
where 
\begin{equation}
 \hat F_{\mu\nu} = \partial_\mu\hat A_\nu - \partial_\nu\hat A_\mu 
  - i\hat A_\mu\star \hat A_\nu + i \hat A_\nu \star \hat A_\mu\;, 
\end{equation}
and $\hat A_\mu$ is a $U(1)$ gauge field. Note that $\hat A_\mu$ is Hermitian.
The noncommutative gauge transformation is given by 
\begin{equation}
 \hat\delta_{\hat\lambda}\hat A_\mu = 
 \partial_\mu\hat\lambda - i\hat A_\mu \star \hat\lambda 
 + i\hat\lambda \star \hat A_\mu \equiv \hat D_\mu\hat\lambda\;,
\end{equation}
with infinitesimal $\hat \lambda = \hat\lambda^*$. Seiberg and Witten 
have shown that an expansion in $\theta$ leads to a map between the noncommutative 
gauge field $\hat A_\mu$ and the commutative gauge field $A_\mu$  as 
well as their respective gauge parameters $\hat \lambda $
and $\lambda$, known as the Seiberg-Witten (SW) map:
\begin{eqnarray}
 \hat A_\mu (A) &=& A_\mu - \frac{1}{2}\theta^{\rho\sigma}A_\rho(\partial_\sigma A_\mu + F_{\sigma\mu})
 + O(\theta^2)\\
 \hat\lambda (\lambda, A) &=& \lambda - \frac{1}{2}\theta^{\rho\sigma}A_\rho\partial_\sigma\lambda + O(\theta^2),
\end{eqnarray}
where the Abelian field strength is given by 
\begin{equation}
 F_{\mu\nu} = \partial_\mu A_\nu - \partial_\nu A_\mu .
\end{equation}
In noncommutative quantum mechanics are often called or referred to as the 
Seiberg -Witten (SW) maps 
linear (non canonical) transformations:
\begin{equation}
 \hat x_i = \hat x_i (x_j, p_j) \quad \hat p_i = \hat p_i (x_j, p_j)
\end{equation}
that relate the extended Heisenberg Algebras in equation (\ref{eq1}) or in equation (\ref{eq2}) to the standard 
Heisenberg algebra in equation (\ref{eq3}).  The Seiberg Witten (SW) map is not unique \cite{bertolami}.
Using one of these transformations, that is a particular 
SW map, it is possible to 
find a representation of the noncommutative observables as operators acting on the conventional 
Hilbert space of ordinary quantum mechanics. The states of the system are then wave functions of the 
ordinary Hilbert space, the Hamiltonian depend on the noncommutativity parameter in the Schro\"dinger 
equation. The Seiberg-Witten (SW) map is considered as a variant of the Bopp's shift.     

{\it Reference papers} :
\item  J. Gamboa, M. Loewe, F. Mendez, J. C. Rojas, Noncommutative quantum mechanics: 
The two-dimensional central field, Int. J. Mod. Phys. A17 2555-2566, (2002);
\item  Akira Kokado, Takashi Okamura, Takesi Saito, 
Noncommutative quantum mechanics and the Seiberg-Witten map, Phys.Rev. D69 (2004) 125007;
\item  Catarina Bastos, Orfeu Bertolami, Nuno Dias, Joao Nuno Prata,
Noncommutative Quantum Mechanics and Quantum Cosmology, Int. J. Mod. Phys. A24 2741-2752, (2009);
\item  O. Bertolami, P. Leal, Aspects of phase-space noncommutative quantum mechanics, Phys. Lett. B,
Volume 750, 12 (2015).
 
\subsection{Path integral formulation }

We consider the following deformed Heisenberg algebra
\begin{equation}\label{pha}
 [\hat q_1,\;\hat q_2] = i\theta;\quad [\hat q_1,\;\hat p_1] = i\hbar;\quad [\hat q_2,\;\hat p_2] = i\hbar;
 \quad [\hat p_1,\;\hat p_2] = i\eta,\;[\hat q_2,\; \hat p_1] = 0,\; [\hat q_1,\; \hat p_2] = 0\;.
\end{equation}
The aim is to provide a path integral formulation of quantum dynamics which is consistent with equation (\ref{pha}). 
A phase space path integral is then formulated from which the commutation relations in equation (\ref{pha}) and 
the extended Heisenberg equations of motion are derived. This is formulated as follows.

In $(2 + 1)$-dimensional space-time, we consider the classical action 
\begin{equation}\label{pathact}
 S  = \int_0^T dt \left(\frac{1}{2}\omega_{ij}x_i\dot x_j - H(x)\right), \quad x_{1,2,3,4} = q_1,q_2,p_1,p_2, 
\end{equation}
where $H$ is the hamiltonian of the system and $\omega_{ij} = (\Theta^{-1})_{ij}$ with 
\begin{equation}
 \Theta = \left(
 \begin{array}{cccc}
  0 & \theta & 1 & 0\\
  -\theta & 0 & 0 & 1\\
  -1 & 0 & 0 & \eta\\
  0 & -1 & -\eta & 0
 \end{array} \right) ; \quad 
  \omega = \frac{1}{1-\theta\eta}\left(\begin{array}{cccc}
  0 & \eta & -1 & 0\\
  -\eta & 0 & 0& -1 \\
  1 & 0 & 0 & \theta \\
  0 & 1 & -\theta & 0
 \end{array}\right).
\end{equation}
We assume here that $\hbar = 1$, and that the matrix $\Theta$ is non singular, means $\theta\eta\neq 1$.
The Hamiltonian equations of motion and the basic Poisson brackets are respectively
\begin{equation}
 \dot x_i = \{x_i, H\} = \Theta_{ij}\frac{\partial H}{\partial x_j},\:\textrm{and}\: \quad \{x_i,x_j \} = \Theta_{ij}.
\end{equation}
A phase space path integral formulation of the quantum theory corresponding to action (\ref{pathact}) is 
\begin{equation}\label{pathint}
 Z = \int \prod_{ k =1}^4 \mathcal{D}x_k e^{iS} = \int \prod_{k =1}^4\mathcal{D}x_k e^{i\int dt (\frac{1}{2}\omega_{ij}x_i\dot x_j - H(x))}.
\end{equation}
\begin{itemize}
\item $Z$ represents a transition amplitude between two states of a given Hilbert space, 
\item The time ordering of operators is enforced, as usual, by the path integral , 
\begin{equation}
\int \mathcal{D}x O_1O_2 e^{iS} = \langle T\{\hat O_1\hat O_2\}\rangle.
\end{equation}
\item Through discretization of the path integral (\ref{pathint}) and using the time 
ordering of operators one derives the commutations relations 
\begin{equation}
[\hat x_i,\hat x_j] = i\Theta_{ij} = i(\omega^{-1})_{ij}
\end{equation}
and the extended Heisenberg equations of motion
\begin{equation}
 \frac{d}{dt}\hat x_i = \Theta_{ij}\frac{\partial \hat H}{\partial \hat x_j} = -i[\hat x_i, \hat H].
\end{equation}
\end{itemize}

Starting with the commutations relations (\ref{pha}), the path integral (\ref{pathint}) 
has been derived showing then the equivalence between the path integral and the operatorial
formalisms. For the later formalism, a Schr\"odinger formulation have been performed as follows.

From the algebra (\ref{pha}) we have  $[\hat q_2,\; \hat p_1] = 0,\; [\hat q_1,\; \hat p_2] = 0$ 
and that provide a basis, for instance, by the set of eivenvectors of $\hat q_1$ and $\hat p_2$, 
$\{\vert q_1,p_2\rangle\}$, or alternatively $\{\vert q_2, p_1\rangle\}$. For an arbitrary state 
$\vert \psi \rangle$, define the wave function (half in coordinate space, half in momentum space) 
$\psi(q_1,p_2,t)\equiv \langle \psi(t)\vert q_1,p_2\rangle$. The action of the operators $\hat q_2$ 
and $\hat p_1$ on the wave function $\psi$ are respectively :
\begin{eqnarray}
 \hat q_2 \psi &=& i (\partial_{p_2} -\theta\partial_{q_1)}\psi ; \\
 \hat p_1 \psi &=& i (-\partial_{q_1} + \sigma\partial_{p_2)}\psi .
\end{eqnarray}
Using the above settings the transition amplitude is calculated and its path integral expression
is derived.

Some comments about the path integral formulation are the following:
\begin{itemize}
 \item The formulation has been done in $(2+1)$ dimensional space-time, considerations 
 can be easily extended to higher dimensional spaces.
 \item The additional quadratic couplings among phase space variables would not complicate substantially 
 the evaluation of a noncommutative partition function, once the commutative case is under control.
 \item When the matrix $\Theta$ is singular, the initial two-dimensional 
 problem reduces to the one dimensional one.
\end{itemize}
Some attempts to introduce path integrals in noncommutative quantum mechanics 
have been discussed in \cite{patha,chachaa,pathb,smailagicac, pathc,scholtzd,pathd}.

{\it Reference papers}:
\item  Ciprian Acatrinei, Path Integral Formulation of Noncommutative Quantum Mechanics,  
JHEP 0109 (2001) 007.
\item Ciprian Acatrinei, Lagrangian versus Quantization, Journal of Phys. A: Math-Gen vol: 37 issue 4 (2004).

\subsection{Weyl-Wigner formulation (phase space)}

Let's consider in a d-dimensional space with noncommuting position and momentum variables, 
the extended Heisenberg algebra reads:
\begin{equation}\label{eha}
 [\hat q_i,\hat q_j] = i\theta_{ij};\quad 
 [\hat q_i, \hat p_j] = i\hbar\delta_{ij}, \quad 
 [\hat p_i,\hat p_j] = i\eta_{ij},\quad, i,j =1,\ldots, d,
\end{equation}
where $\eta_{ij}$ and $\theta_{ij}$ are antisymmetric real constant $(d\times d)$ matrices 
and $\delta_{ij}$ is the identity matrix. We assume that 
\begin{equation}
 \Sigma_{ij} \equiv \delta_{ij} + \frac{1}{\hbar^2}\theta_{ik}\eta_{kj} 
\end{equation}
is equally an invertible matrix. That means that for any matrix elements $\eta$ and $\theta$, 
their product is considerably smaller than $\hbar^2$: 
\begin{equation}
 \theta\eta < < \hbar^2.
\end{equation}

The algebra (\ref{eha}) is related to the standard Heisenberg algebra 
\begin{equation}
 [q_i, q_j] = 0,\quad 
 [q_i, p_j] = i\hbar\delta_{ij},\quad 
 [p_i, p_j] = 0,\quad i,j =1,\ldots d, 
\end{equation}
via the Seiberg Witten (SW) map as follows
\begin{equation}\label{trasf}
 \hat q_i =  A_{ij} q_j + B_{ij} p_j, \quad \hat p_i = C_{ij} q_j + D_{ij}p_j,
\end{equation}
where $\bf A,B,C,D$ are real constant matrices. The transformation (\ref{trasf}) 
is assumed to be inversible.

In order to present the Weyl-Wigner formulation of noncommutative quantum mechanics, let's recall briefly the 
following.

The non-covariant Weyl-Wigner map is an isomorphism between the operator and the phase space representations 
of the ordinary quantum mechanics based on the standard Heisenberg algebra providing the simplest approach 
to derive most of the mathematical structure of conventional phase space quantum mechanics \cite{groe,wey, wig}.
The covariant  extension of this map was studied in \cite{nuno} in connection with a diffeomorphism invariant 
formulation of Weyl-Wigner quantum mechanics.  

The Weyl-Wigner formulation  of noncommutative quantum mechanics relies on 
the covariant generalization of the Weyl-Wigner transform 
and the Seiberg-Witten (SW) map. The extended Weyl-Wigner map for noncommutative quantum mechanics is constructed.
It is an isomorphism between the operator and phase space representations of the extended 
algebra (\ref{eha}), providing  a systematic approach to derive the phase space formulation of noncommutative 
quantum mechanics in its most general form. 
  It has been shown that the noncommutative Weyl-Wigner transform, and therefore the entire formulation of 
noncommutative quantum mechanique does not depend on the particular choice for the SW map.
\begin{itemize}
 \item The noncommutative Wigner function 
 is obtained by applying the extended Weyl-Wigner transform to the density matrix. 
 It is the noncommutative counterpart of the Wigner function.
 \item The extended $\star$-product and an extended  Moyal bracket are derived.
 \item The dynamical and eigenvalue equations for noncommutative quantum mechanics are given.
\end{itemize}
Details about this approach are found in the reference paper :
\item  Catarina Bastos, Orfeu Bertolami, Nuno Costa Dias, Joao Nuno Prata,
Weyl-Wigner Formulation of non-commutative Quantum Mechanics, 
J. Math. Phys. 49:072101, 2008.

Further development of the current formulation where a number of issues related to the characterization 
of the set of states of the theory is adressed in the following paper:
\item  C. Bastos, N. C. Dias, and J. N. Prata,  Wigner Measures in noncommutative quantum mechanics, 
 Comm. Math. Phys. 299, 709-740, 2010.

\subsection{Systematic formulation }\label{sec4}
 
We start by brieffly introducing the noncommutative analog of field derivatives \cite{chaichian}.
In the noncommutative approach, we consider the Hermitian operators (positions) 
$\hat x_i, \; (i = 1,2)$ satisfying the commutation relations 
\begin{equation}
 [\hat x_i,\hat x_j] = i\lambda^2 \epsilon_{ij},\quad i,j =1,2,
\end{equation}
where $\lambda$ is a positive constant of the dimension of length.
The noncommutative analogs of field derivatives 
are defined as follows
\begin{equation}\label{derv}
 \partial_i\hat\varphi = \epsilon_{ij}\frac{i}{\lambda^2}[\hat x_j,\;\hat\varphi] \quad i = 1,2.
\end{equation}
They satisfy the Leiniz rule and reduce to the usual derivatives of the commutative limit. 
Using this definition (\ref{derv}) the momenta operators should be 
\begin{equation}
 \hat p_i = i\hbar\lambda^{-2}\epsilon_{ij}ad_{x_j},
\end{equation}
$ad_{x_i}\hat A \equiv [\hat x_i, \hat A]$ for any operator $\hat A$. It is easy to verify that
 $[\hat p_1,\;\hat p_2] = 0$.
 
In this section we present a formalism of noncommutative quantum mechanics in 
complete analogy with conventional quantum mechanics as a quantum system on the Hilbert space 
of Hilbert Schmidt operators acting on noncommutative configuration space.

In two dimensions, the coordinates of non-commutative configuration space satisfy the commutation relation 
\begin{equation}\label{equo1}
 [\hat x_i, \hat x_j] = i\theta\epsilon_{ij},
\end{equation}
with $\theta$ being a real positive parameter and $\epsilon_{ij}$ the completely anti-symmetric tensor 
with $\epsilon_{1,2} = 1$. From (\ref{equo1}), we define creation and annihilation operators 
\begin{equation}
 b = \frac{1}{\sqrt{2\theta}}(\hat x_1 + i\hat x_2), \quad 
 b^\dagger = \frac{1}{\sqrt{2\theta}}(\hat x_1 - i\hat x_2)
\end{equation}
that satisfy the Fock algebra $[b, b^\dagger] = 1$, the noncommutative configuration space is 
isomorphic to boson Fock sapce
\begin{equation}
 \mathcal{H}_c = \textrm{span}\lbrace\vert n \rangle \equiv \frac{1}{\sqrt{n!}}(b^\dagger)^n\vert 0\rangle\}_{n=0}^{n=\infty}
\end{equation}
where the span is taken over the field of complex numbers.

We consider now the set of Hilbert-Schmidt operators acting on the noncommutative configuration space 
\begin{equation}
 \mathcal{H}_q = \{ \psi(\hat x_1,\hat x_2): \psi(\hat x_1,\hat x_2)\in \mathcal{B}(\mathcal{H}_c), 
 \textrm{tr}_c(\psi(\hat x_1,\hat x_2)^\dagger\psi(\hat x_1,\hat x_2)) < \infty \},
\end{equation}
where $\textrm{tr}_c$ denotes the trace over non-commutative configuration space, 
$\mathcal{B}(\mathcal{H}_c)$ denotes the set of bounded operators on $\mathcal{H}_c$.
In other words, the Hilbert space is the trace class enveloping algebra of the classical configuration 
space Fock algebra $(b,b^\dagger)$. As these operators are necessarily bounded, this is again a Hilbert 
space (recall that the set of bounded operators on a Hilbert space is again a Hilbert space) and to 
distinguish the classical configuration space, which is also a Hilbert space, from the quantum Hilbert space 
we use, respectively, c and q as subscripts. We follow the same notation to distinguish operators acting on the 
classical or quantum Hilbert space. Furthermore we denote states in the quantum Hilbert space by $| \cdot )$ 
and states in the classical configuration space by $| \cdot \rangle$. The corresponding inner product is $(\psi\vert \phi)
 = (\psi,\phi) = \textrm{tr}_c(\psi^\dagger\phi)$, which also serves to define bra states as elements of the 
 dual space (linear functionals). Note that the trace is performed over the classical 
 configuration space, denoted by subscript c. In order to distinguish the notations for Hermitian conjugation 
 between the two Hilbert spaces, we use the notation $\dagger$ to denote Hermitian conjugation on noncommutative 
 configuration space and the notation $\ddagger$ for Hermitian conjugation on quantum Hilbert space.

The abstract Heisenberg algebra is now replaced by the non-commutative Heisenberg algebra.
In two dimensions, we have 
\begin{eqnarray}\nonumber 
{[x_i,p_j ]}  &=&  i\hbar \delta_{ij},\\\label{anhe}
{[x_i,x_j ]}  &=&  i\theta\epsilon_{ij},\\\nonumber
{[p_i,p_j ]}  &= & 0.
\end{eqnarray}

A unitary representation of this algebra in terms of operators $\hat X_i$ and $\hat P_i$ 
acting on the quantum Hilbert space $\mathcal{H}_q$ which is the analog of the Schr\"odinger 
representation of the Heisenberg algebra is 
\begin{eqnarray}\nonumber
 \hat X_i \psi(\hat x_1,\hat x_2) &=& \hat x_i\psi(\hat x_1,\hat x_2),\\\label{srep}
 \hat P_i \psi (\hat x_1,\hat x_2) &=& \frac{\hbar}{\theta}\epsilon_{ij}[\hat x_j, \psi(\hat x_1,\hat x_2)],
\end{eqnarray}
the position operator acts by lelft multiplication and the momentum operator adjointly. 

In the equation (\ref{anhe}), we have considered a situation where only the coordinates are 
noncommutative. This was necessary to write down the representation (\ref{srep}) which 
requires commuting momenta to be consistent. We consider now
systems in which the momenta are also noncommutative, for instance in the presence of magnetic field,  where the 
set of commutations relations are 
\begin{eqnarray}\nonumber
{[ x_i,p_j ]} & = & i\hbar \delta_{ij},\\\label{manhe}
{[ x_i,x_j ]} &=& i\theta\epsilon_{ij},\\\nonumber
{[p_i,p_j ]} &=& i\gamma\epsilon_{ij},
\end{eqnarray}
In order to apply our formalism, we can bring these commutations 
relations in equation (\ref{manhe}) in the same form as (\ref{anhe}) through an appropriate linear transformation on 
momenta and coordinates. 
We consider a transformation to new coordinates $y_i$ and momenta $\pi_i$
given by 
\begin{equation}
 \left\{ 
 \begin{array}{cc}
 y_i   = &  x_i \\
 \pi_i =  & \alpha p_i + \beta\epsilon_{ij}x_j
 \end{array}\right.
\end{equation}
The new coordinates and momenta satisfy the equation (\ref{anhe}) with the following choice 
of $\alpha$ and $\beta$ 
\begin{equation}
 \alpha = \frac{\pm\hbar}{\sqrt{\hbar^2 -\gamma\theta}},\quad \beta = \frac{\hbar}{\theta}(1-\alpha).
\end{equation}
The formalism can be applied now to the new coordinates, while maintaining the potential 
as a function of coordinates only. There is a critical value of the parameter $\gamma = \frac{\hbar^2}{\theta}$
such that for $\gamma < \frac{\hbar^2}{\theta}$ this representation is unitary but for $\gamma > \frac{\hbar^2}{\theta}$
it loses its unitarity.

With the notions that we set, one can proceed with the normal quantum mechanical interpretation in 
the quantum Hilbert space $\mathcal{H}_q$. 

{\it Reference papers} :

\item Formulation, Interpretation and Application of non-commutative Quantum Mechanics, 
F. G. Scholtz, L. Gouba, A. Hafver, C. M. Rohwer, J. Phys. A 42:175303, 2009.

The present formulation has been applied succesfully to much more difficult potentials such as
the spherical well. A precise meaning to piecewise constant potential in noncommutative quantum me-
chanics is given using this formulation. More details can be found in the following paper:
\item F. G. Scholtz, B. Chakraborty, J. Govaerts, S. Vaidya, Spectrum of the non-commutative spherical well, 
 J. Phys. A Math-theor. 40.  Issue: 48 Pages: 14581-14592, 2007.

An unambiguous formulation of the path integral representation for the propagator has been 
presented using this formulation. An action for a particle moving in the non commutative plane and
in the presence of an arbitrary potential is derived. More details can be found in the following paper:
\item S. Gangopadhyay, F. G. Scholtz, Path-Integral Action of a Particle in the Noncommutative Plane, 
PRL Volume: 102. Issue: 24  Article Number: 241602, 2009.

The Gazeau-Klauder coherent states in noncommutative quantum mechanics have been studied using the 
this formulation. The inherent vector feature of these states has been revealed. Details are found in:
 \item J. Ben Geloun and F. G. Scholtz, Coherent states in noncommutative quantum mechanics, 
 J. Math. Phys. 50, 043505, 2009.
 
This formulation has been used to present a description of noncommutative
quantum mechanics which may be viewed in terms of spatially extended objects. More 
details about this description is found in the following paper:
\item C. M. Rohwer, K. G. Zloshchastiev, L. Gouba, F. G. Scholtz,
Noncommutative quantum mechanics-a perspective on structure and spatial extent, 
J. Phys. A Math-Theor. {\bf 43}   Issue: 34, 2010.

\section{Concluding remarks}

Noncommutative quantum mechanics represents a natural extension of usual quantum mechanics, 
in which one allows nonvanishing commutators also between the coordinates and between the momenta in general. 
The notion of coordinates basis and the very concept of wave functions $\langle x,\psi\rangle$ 
fails. However the usual momentum space description is still valid in case the momenta commute 
but in the general case where the momenta do not commute it does not hold. 

Various formulations of noncommutative quantum mechanics have been constructed and we attempt to list them
in this paper. 

For the canonical formulation that assemble the new coordinates system, the Bopp's shift method, 
the star product method and the Seiberg Witten map,  the procedure is to relate the extended Heisenberg algebra 
to the standard Heisenberg algebra by a class of linear transformations. The Bopp's shift method is equivalent 
to the star product method and it is considered as an older version of the Seiberg-Witten map. The problem with the 
canonical formulation is that both the observables and the states beings written in terms of 
the Heisenberg variables do not display a simple mathematical structure. This tends to obscure the 
physical meaning. Shifting the potential involves in principle arbitrary large number of derivations 
in the Schr\"odinger equation.  

For the path integral formulation, the procedure is to provide a phase space path integral, 
starting from a classical action, that is consistent with the deformed (extended) Heisenberg algebra. 
The equivalence between the path integral and the operatorial formulation has been proved. 
Despite the claims that noncommutativity of coordinates may bar having a Lagrangian description \cite{acatri}, 
there exits such a formulation in \cite{chachaa},
where an effective Lagrangian has been derived from 
a path integral approach and quasi-classical approximation. 

For the Weyl-Wigner formulation, the procedure is to use the Seiberg-Witten map and the 
covariant generalization of the Weyl-Wigner transform to construct an isomorphism 
between the operator and the phase space representations of the extended Heisenberg algebra. 
This formulation is useful for treating general problems such as, for instance, in 
case where the potential is not specified.

For the systematic formulation, the procedure is to formulate noncommutative quantum 
mechanics, in complete analogy with commutative quantum mechanics, as a quantum system 
on the Hilbert space of Hilbert-Schmidt operators acting on noncommutative configuration 
space. This approach presents a consistent formulation and interpretational framework for 
non-commutative quantum mechanics, which includes an unambiguous description for position 
measurement. 

Revising the four formulations, the path integral and the systematic approaches do not
use a change of coordinates ($\star$-product, Bopp's shift, SW maps). 
However, in a general setting where in addition the momenta do not commute we have the following 
situations: the path integral approach still holds in case there are some commuting phase space coordinates left 
so that a complete basis in the Hilbert space of the theory is provided by the set of the eigenvectors of the 
commuting phase space coordinates; for the systematic approach appropriate linear transformations on momenta 
and coordinates are needed. 

\end{myenumerate}

{\bf Acknowledgments}

I would like to gratefully thank  M. M. Sheikh-Jabbari for constructive discussions.
I would also like to thank Peter Horvathy, Subir Ghosh, Ali Alavi, Alexei Deriglazov, Mikhail Plyushchay for 
comments on the manuscript. Finally, I gratefully acknowledge support from 
the Abdus Salam International Centre for Theoretical Physics (ICTP), Trieste, Italy.

\end{document}